\documentclass[runningheads]{llncs}
\usepackage[hyphens]{url}

\usepackage{graphicx}
\begin{document}
\title{On-Demand and Lightweight Knowledge Graph Generation -- a Demonstration with DBpedia}
\titlerunning{On-Demand and Lightweight Knowledge Graph Generation}
%
\author{Malte Brockmeier \and Yawen Liu \and Sunita Pateer \and\\ Sven Hertling\orcidID{0000-0003-0333-5888} \and Heiko Paulheim\orcidID{0000-0003-4386-8195}}
\authorrunning{M. Brockmeier et al.}
%
\institute{Data and Web Science Group\\University of Mannheim, Germany\\
\email{\{mbrockme,yawliu,spateer\}@mail.uni-mannheim.de}\\
\email{\{sven,heiko\}@informatik.uni-mannheim.de}\\
\url{http://dws.informatik.uni-mannheim.de/}}
\maketitle              
\begin{abstract}
Modern large-scale knowledge graphs, such as DBpedia, are datasets which require large computational resources to serve and process. 
Moreover, they often have longer release cycles, which leads to outdated information in those graphs. 
In this paper, we present \emph{DBpedia on Demand} -- a system which serves DBpedia resources on demand without the need to materialize and store the entire graph, and which even provides limited querying functionality. 
\keywords{Knowledge Graph \and On Demand \and Live \and Lightweight \and Knowledge Graph as a Service \and DBpedia}
\end{abstract}

\section{Introduction}
Knowledge graphs on the Web -- such as DBpedia \cite{lehmann2015dbpedia}, YAGO \cite{rebele2016yago}, or Wikidata \cite{vrandevcic2014wikidata} -- are useful building blocks for intelligent applications \cite{heist2020knowledge}. While they differ considerably in coverage and quality \cite{farber2018linked}, in many cases, timeliness of information is a crucial property in some use cases as well. Since many public knowledge graphs have rather long release cycles, sometimes up to a few months or even years \cite{heist2020knowledge}, the information served in these knowledge graphs may be easily outdated. One approach which has been proposed to close this gap is \emph{DBpedia Live}, which provides fresh data extracted from Wikipedia as a knowledge graph \cite{morsey2012dbpedia}, and which, after a longer downtime, is available again since 2019.\footnote{\url{https://www.dbpedia.org/blog/dbpedia-live-restart-getting-things-done/}}

Another challenge related to such knowledge graphs is their sheer size. The creation of those graphs is often a longer running and computation-intensive process, and even serving the knowledge graph online requires quite a bit of computing power. The hosted services for hosting a static DBpedia copy or running DBpedia live cost around \$1 per hour at the time of writing this paper.\footnote{\url{https://aws.amazon.com/marketplace/seller-profile?id=1b9b499d-15e3-479d-a644-da16d45c40a7}}

In this paper, we introduce \emph{DBpedia on Demand} -- an installation of DBpedia which can be run on a local machine as a service. It creates knowledge graph resources on request without materializing an entire graph, which makes its computing requirements rather minimal. It provides a Linked Data endpoint as well as limited SPARQL querying properties.\footnote{Code available at \url{https://github.com/mbrockmeier/KnowledgeGraphOnDemand}}

\section{Prototype}
\begin{figure}[t]
    \centering
    \includegraphics[width=\textwidth]{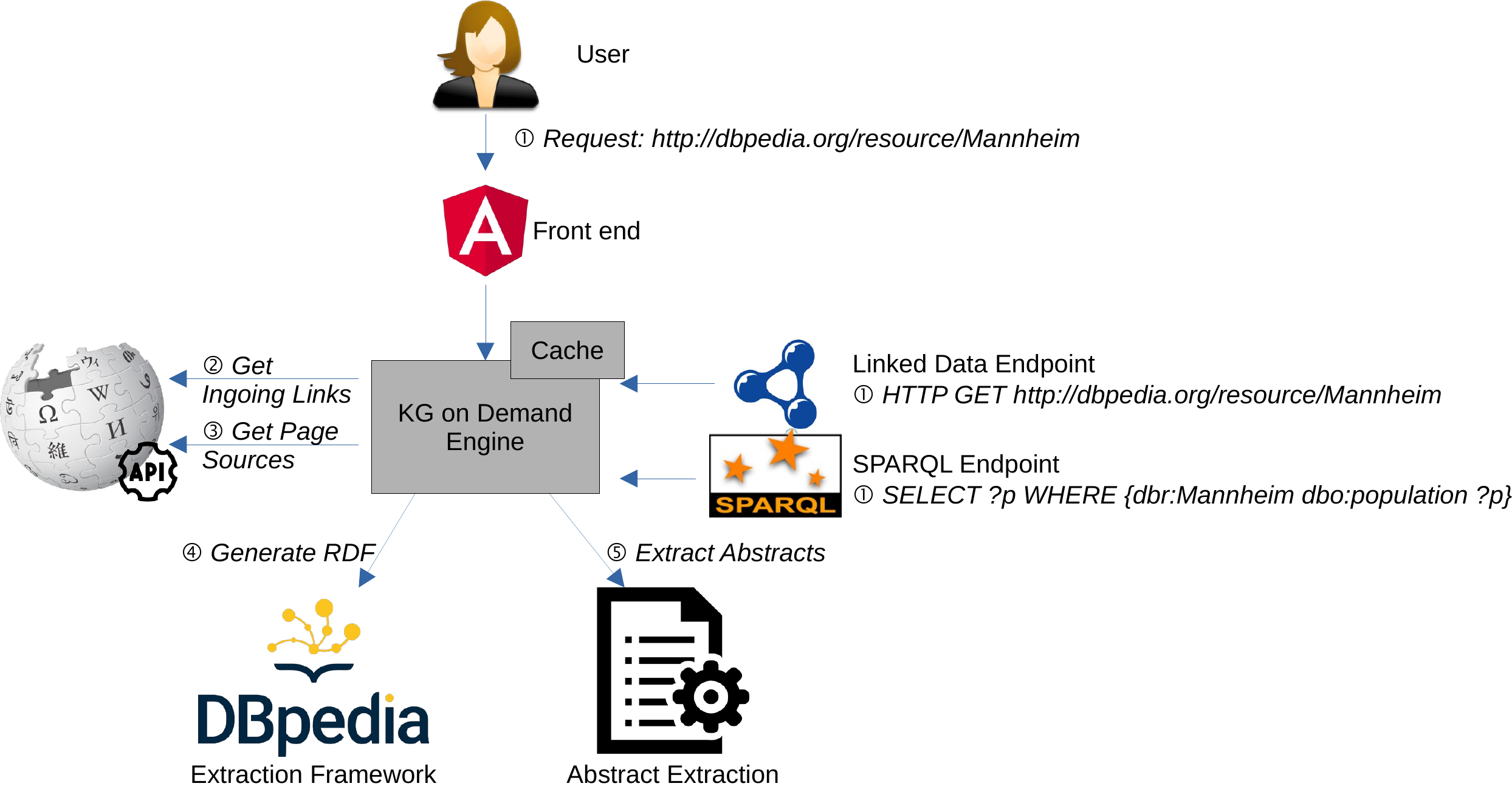}
    \caption{Overall System Architecture of DBpedia on Demand}
    \label{fig:architecture}
\end{figure}
DBpedia is built from Wikipedia by extracting information from infoboxes and other bits of structured contents, such as categories. To that end, it uses \emph{mappings} from infoboxes to a central ontology, which are also stored in a Wiki. When creating a release, a dump of Wikipedia is processed, and the page contents are transformed to knowledge graph instances using the mappings defined in the Mappings Wiki \cite{lehmann2015dbpedia}.

The idea of KG on demand is different. Instead of creating a full DBpedia copy, it creates instances upon request: when a user requests a certain resource, the corresponding Wikipedia page is downloaded and processed in real time. Mappings can be downloaded from the mappings Wiki regularly, which allows for using at least recent, although not up to date mappings. Fig.~\ref{fig:architecture} shows the overall architecture of the KG on Demand engine. The process runs in five steps:
\begin{enumerate}
    \item The URI for which to generate the RDF for is obtained, either trough the user interface, a HTTP request, or by extracting it from a SPARQL query.
    \item The ingoing links (i.e., Wikipedia pages pointing to the page which corresponds to the requested resource) to that resource are obtained from the Wikipedia API.
    \item The page sources for both the requested resources and the resources of the pages linking to the resource are obtained via the Wikipedia API.
    \item For all those pages, the corresponding RDF representation is generated via the DBpedia Extraction Framework.
    \item Additionally, the abstract for the resource at hand is generated by extracting the first few sentences from the Wikipedia page.
\end{enumerate}
In the DBpedia knowledge graph, outgoing edges from a resource are generated from the corresponding Wikipedia page's infobox. Hence, for generating only the outgoing edges, processing one single Wikipedia page is sufficient. In contrast, ingoing edges are outgoing edges of \emph{other} Wikipedia pages. In order to generate those, we first query the Wikipedia API for other pages that link to the one at hand, and then process those as well in order to generate the ingoing edges.

Since the abstract extraction in DBpedia requires the creation and processing of a local copy of the entire Wikipedia, we developed or own simple abstract extraction mechanism instead of using the original one.

\begin{figure}[t]
    \centering
    \includegraphics[width=\textwidth]{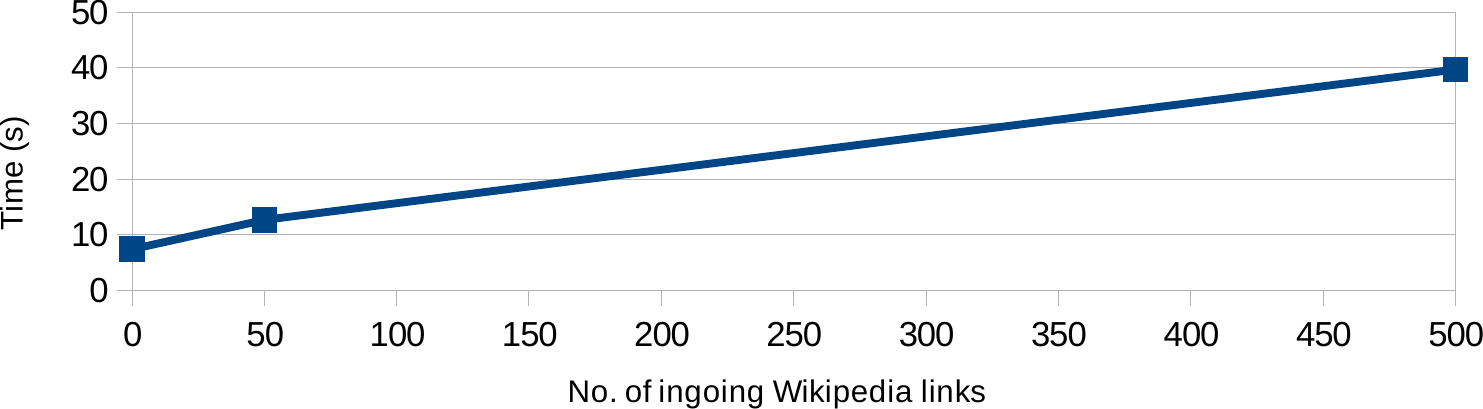}
    \caption{Performance of Knowledge Graph on Demand}
    \label{fig:performance}
\end{figure}

The number of ingoing Wikipedia links is the key factor that influences the performance of the proposed approach. Fig.~\ref{fig:performance} depicts the performance for extracting resources from Wikipedia pages.\footnote{The tests were run on an Intel Core i7-8700K processor, 32GB of DDR4 RAM, a 1TB NVMe SSD, and an internet connection with a downstream rate of 100MBit/s. The reported processing times are the averages of 10 runs.}
It can be observed that the systems scales linearly with the number of ingoing links.\footnote{Since the number of ingoing links are the main driver of performance, the maximum number of ingoing links to consider can also be reduced in the configuration of our implementation. By default, all ingoing links are processed.}

The approach of generating resources for URIs also determines the kind of SPARQL queries that our approach can process. Generally, all queries that only consider simple ingoing and outgoing edges of a fixed resource can be handled, such as
\begin{verbatim}
SELECT ?actor WHERE { ?actor dbo:starring dbr:Lost_Highway .}
SELECT ?director WHERE { dbr:Lost_Highway dbo:director ?director .}
\end{verbatim}
On the other hand, our approach cannot handle queries without a fixed resource, or queries with a path length greater than one, such as:
\begin{verbatim}
SELECT ?actor ?movie WHERE { ?actor dbo:starring ?movie .}
SELECT ?director WHERE { dbr:Tom_Cruise dbo:starring ?movie .
                         ?movie dbo:director ?director .}
\end{verbatim}
In other words, the queries are limited to patterns with at least one resource fixed, and all query variables being at most one hop away from that resource.

\section{DBpedia Live vs. DBpedia on Demand}
DBpedia live builds a full copy of DBpedia. This copy is then updated by monitoring the update stream of Wikipedia. Whenever a change is made in Wikipedia, DBpedia live is notified and processes the change by re-computing the corresponding instance representation in the knowledge graph. Likewise, it is notified on changes in the Mappings wiki (e.g., if the mapping of an infobox to an ontology class is changed), and the affected instances can then be changed as well.

By design, DBpedia live always stores a local materialized copy of the full knowledge graph, which leads to high computational costs of serving the knowledge graph endpoint (i.e., it can be run on commodity hardware). Since all requests are processed on that local copy, they can be handled faster than in DBpedia on demand, and also all types of SPARQL queries can be answered. At the same time, DBpedia Live requires a lot of online runtime to constantly process the stream of changes from Wikipedia.
Table~\ref{tab:live_vs_ondemand} summarizes the main differences between DBpedia Live and DBpedia on Demand.

\begin{table}[t]
    \caption{Comparison of DBpedia Live and DBpedia on Demand}
    \label{tab:live_vs_ondemand}
    \centering
    \begin{tabular}{r|l|l}
         & DBpedia Live & DBpedia on Demand\\
         \hline
         Storage requirements & high & low \\
         \hline
         Online runtime & high & low \\
         \hline
         Response time & fast & slower \\
         \hline
         Latest mapping version & yes & approximate \\
         \hline
         SPARQL interface & full & limited \\
    \end{tabular}
\end{table}

\section{Conclusion and Outlook}
In this paper, we have shown an approach which generates a knowledge graph like DBpedia in an on-demand and lightweight fashion. The approach is very resource-efficient and can be run on commodity hardware. Hence, it is an interesting building block for applications which use a knowledge graph, but do neither want to rely on a public endpoint, nor materialize an entire knowledge graph. The approach can be seen as a complement to DBpedia Live, in comparison to which it has different advantages and disadvantages.

While the current implementation is a prototype only, there are some shortcomings which we have inherited from the DBpedia Extraction Framework. First, that framework is not capable of multithreading, which makes the current implementation suitable for a local service, rather than for setting up a public endpoint. Second, we have currently encapsulated the \emph{entire} extraction framework, which, however, comes with a significant ramp-up time. Both issues could be addressed by branching and refactoring the extraction framework's codebase.

While this demonstration has been based on DBpedia, it can be transferred to other approaches as well. With the same mechanism, it would be possible to use the extraction code of other Wikipedia-based knowledge graphs, such as YAGO~\cite{rebele2016yago} or the Linked Hypernym extraction~\cite{kliegr2015linked}, as well as to transfer the approach to other Wikis~\cite{hertling2020dbkwik}. Also other refinement operators which are local to a Wikipedia page, such as extracting relations from text~\cite{heist2018language}, would be applicable.

While the SPARQL capabilities in our approach are still basic, the approach could be extended towards supporting more complex queries. Possible approaches would be incremental solving of queries, or first using the Wikipedia link graph for obtaining candidates, and then running the extraction on the candidates to narrow down the solution set. 
For implementing a query endpoint, it would also be interesting to build a (possibly limited) linked data fragments endpoint \cite{verborgh2016triple} with the approach at hand, since many of the basic building blocks of those endpoints (e.g., paging) can be directly transferred to our approach.

Apart from adding querying capabilities, the endpoint could also be enriched with an on-demand approach of generation of \emph{embedding vectors}, such as \emph{RDF2vec Light} \cite{portisch2020rdf2vec}. That way, downstream applications could leverage both explicit as well as latent representations of the entities in the knowledge graph.


\bibliographystyle{splncs04}
\bibliography{references}
\end{document}